\documentclass[final,5p,fleqn]{elsarticle}
\usepackage{graphicx}
\usepackage{hyperref}
\usepackage{amsmath}
\usepackage{url}
\usepackage{colordvi}
\usepackage{subfigure}
\usepackage{times}

\begin{document}

\begin{frontmatter}

\title{ 
A New Extraction of Pion Parton Distributions  
in the Statistical Model}

\author[a] {Claude Bourrely}
\author[b] {Franco Buccella}
\author[c] {Jen-Chieh Peng}

\address[a]{Aix Marseille Univ, Universit\'e de Toulon, CNRS, CPT, 
Marseille, France}
\address[b]{INFN, Sezione di Napoli, Via Cintia, Napoli, I-80126, 
Italy} 
\address[c]{Department of Physics, University of Illinois at 
Urbana-Champaign, Urbana, Illinois 61801, USA}

\date{\today}

\begin{abstract}
We present a new analysis to extract pion's parton distribution 
functions (PDFs) in the framework of the statistical model. Starting from
the statistical model framework first developed for the spin-1/2 nucleon, we
apply appropriate modifications taking into account the spin-0 nature of
pion and the isospin and charge-conjugation symmetry properties. 
This results in a significant reduction of the number of parameters  
compared to a recent work to extract pion's PDFs. 
Using $\pi^-$-induced Drell-Yan data to determine the parameters of this
statistical model approach, we show that a good description of these 
experimental data with Next-to-Leading order QCD calculations can be 
obtained. Good agreement between the calculations and the $\pi^+/\pi^-$
Drell-Yan cross section ratio data, not included in the global fit, has
confirmed the predictive power of these new pion PDFs. 
\end{abstract}

\end{frontmatter}

The first determination of the proton parton distribution functions (PDFs),
based on the framework of the statistical model, was proposed about 20 years
ago by Bourrely, Buccella, and Soffer~\cite{BBS2002}. Some salient features
of the statistical approach include the natural connection between the
polarized and unpolarized parton distributions, as well as the relationships
between the valence and sea quark distributions. These important features 
of the statistical approach allow many predictions for the flavor and spin
structures of proton's PDFs, which are usually not possible for
the conventional global fits without  adding theoretical constraints. 
Some notable successes of
the statistical model include the prediction of $x$ distribution of the
flavor asymmetry of unpolarized sea, $\bar d(x) -\bar u(x)$, and the prediction
of $\Delta \bar u(x) > 0 > \Delta \bar d(x)$~\cite{BBS2005}.
These predictions were confirmed in recent
unpolarized Drell-Yan experiment~\cite{Dove2019} and single-spin asymmetry
measurement of W-boson production~\cite{STAR2014}. A review of the major
results on describing the nucleon parton distributions in the statistical
model can be found in~\cite{BS2015}.

The statistical approach for extracting proton's PDFs
can be naturally extended to other hadrons. Of particular interest are the
PDFs for pions. Pion has the dual roles of being the lightest 
quark-antiquark bound state and a Goldstone boson due to the spontaneous
breaking of the chiral symmetry. Many theoretical models have explored the
partonic structures of the pion~\cite{HORN2016}. Recent 
advance~\cite{Ji} in lattice QCD has also
led to the first calculations on the $x$ distribution of the meson PDFs 
in the Large-Momentum Effective Theory (LaMET)~\cite{Zhang2019,Sufian2020}. 
New experimental data relevant to 
the pion PDFs have been collected in the COMPASS experiment with 
pion-induced dimuon production~\cite{Compass2017}. 
The interesting prospects of probing pion's
PDFs with tagged deep-inelastic scattering are being pursued at the Jefferson
Laboratory and considered for the future Electron-Ion 
Collider (EIC)~\cite{EIC}. 
The interest in pion's partonic structure is reflected in several recent 
publications~\cite{JAM,BS2019,Xfitter} where pion's PDFs were extracted 
via global fits to existing data.

The first extraction of the pion's PDFs based on the statistical model
approach was reported in~\cite{BS2019}. 
In this paper we adopt a much simpler parametrization for the functional
forms of the parton distributions 
by imposing some constraints based on symmetry principles.
We first define
the parametrizations of the pion's quark, antiquark and gluon distributions
based on the statistical model. We then show that the existing pion-induced
Drell-Yan data E615~\cite{e615}, NA10~\cite{NA10} and  E326~\cite{e326}
can be very well described with these new pion PDFs. A prediction
for the ratio $\pi^+/\pi^-$ is also shown, followed by conclusion.

In the previous analysis to extract pion's PDFs in the statistical 
model~\cite{BS2019}, no assumption was made on the flavor and spin structures
of the quark and antiquark distributions in pion. While this led to a 
significant flexibility in determining pion's PDFs, the limited amount
of relevant experimental data would limit the
ability to determine these flexible parametrizations in an unambiguous
fashion. It is advantageous to reduce the number of parameters, taking into
account some symmetry principles and other plausible
assumptions. However, it was unclear whether the experimental data can be
well described by the statistical model when the number of parameters is
significantly reduced. The findings of this work, to be discussed later,
are that an excellent description of the data can be obtained in the statistical
model approach when the parameters are reduced in a judicious fashion.

We begin by defining the notations of the various parton distribution functions
for pions. After imposing the particle-antiparticle charge-conjugation (C) 
symmetry for the parton distributions in charged pions, we can define the
PDFs in $\pi^+$ and $\pi^-$ as follows:
\begin{equation}
U(x) \equiv  u_{\pi^+}(x) = \bar u_{\pi^-}(x);~~~D(x) \equiv 
\bar d_{\pi^+}(x) = d_{\pi^-}(x)~.
\label{eq1}
\end{equation}
\begin{equation}
\bar U(x) \equiv \bar u_{\pi^+}(x) = \bar d_{\pi^-}(x);~~~\bar D(x) \equiv
d_{\pi^+}(x) = u_{\pi^-}(x)~.
\label{eq2}
\end{equation}
\begin{equation}
S(x) \equiv s_{\pi^+}(x) = \bar s_{\pi^-}(x);~~~\bar S(x) \equiv 
\bar s_{\pi^+}(x) = s_{\pi^-}(x)~.
\label{eq3}
\end{equation}
\begin{equation}
G(x) \equiv  g_{\pi^+}(x) = g_{\pi^-}(x)~.
\label{eq4}
\end{equation}
In Eqs. (\ref{eq1})-(\ref{eq4}), we define 7 PDFs, namely, $U(x)$,
$D(x)$, $\bar U(x)$, $\bar D(x)$, $S(x)$, $\bar S(x)$ and $G(x)$.
We can further require charge symmetry (CS), which is a weaker form of 
the isospin symmetry, to reduce the number of independent PDFs. 
The CS refers to the invariance under a rotation
by 180$^\circ$ along the second axis in the isospin space. 
We note that
in the previous statistical model analysis~\cite{BS2019}, 
the C symmetry was imposed but the CS was not required.
While it is of great interest
to test the validity of CS in pion PDFs, the existing data are not sensitive
to violation of CS, as discussed in a review of the theories and
experiments on CS at the partonic level~\cite{Tim}. Therefore, the CS is
now required for the present work. This requirement can be relaxed in 
the future when precision data sensitive to CS in pion PDFs become available. 

It is well known that the requirements of C and CS invariance would
imply $U(x) = D(x)$ and $\bar U(x) = \bar D(x)$. As shown in Eq.~(\ref{eq1}),
C symmetry leads to $u_{\pi^+}(x) = \bar u_{\pi^-}(x)$.  Invariance
under the rotation in the isospin space by 180$^\circ$, i.e., CS invariance,
would give $\bar u_{\pi^-}(x) = \bar d_{\pi^+}(x)$. Therefore, invariance
under the combined operations of C and CS implies $U(x) = D(x)$. In a similar
fashion, it can be readily shown that $\bar U(x) = \bar D(x)$.
 
The absence of valence strange quark in pion
only implies that the first moments of $S(x)$ and $\bar S(x)$ are the same,
namely,
\begin{equation}
\int_0^1 [S(x) - \bar S(x)] dx = 0;~~~~\int_0^1 S(x) dx = \int_0^1 \bar S(x) dx.
\label{eq3a}
\end{equation}
Using C and CS invariance, it can be shown that $S(x) = \bar S(x)$ in
Eq.~(\ref{eq3}). First, the C invariance implies 
$s_{\pi^+}(x) = \bar s_{\pi^-}(x)$. A subsequent rotation in the isospin
space would give $\bar s_{\pi^-}(x) = \bar s_{\pi^+}(x)$, because $s$ and 
$\bar s$ are isoscalar particles and invariant under isospin rotation. 
Therefore, we obtain
\begin{equation}
S(x)=\bar S(x)
\label{eq3b}
\end{equation}
as a result of the invariance under a combined operation
of C and CS. In Ref.~\cite{BS2019} the strange quark contents in pion
were ignored. 
In this work, we include
the contributions from strange sea, as they must be present 
and contribute to the momentum sum rule. 

In the statistical approach for proton's PDFs, there are positive and
negative helicity distributions for each quark and antiquark flavor.
Unlike the spin-1/2 proton, pion has zero spin. Hence, it is no longer 
necessary to define the positive and negative helicity distributions
for pion's PDFs. This further simplifies the analysis compared with the
earlier work~\cite{BS2019}, and it is reflected
in Eqs. (\ref{eq1})-(\ref{eq3}) which only define a single parton distribution 
for each quark or antiquark flavor.

Based on the framework of the statistical model, we adopt the following
functional forms for pion's PDFs:

\begin{equation}
xU(x)= xD(x) =\frac{A_U X_U x^{b_U}}{\exp[(x-X_U)/\bar x] + 1}
+\frac{\tilde A_{U} x^{\tilde b_{U}}}{\exp(x/\bar x) + 1}~.
\label{eq5}
\end{equation}

\begin{equation}
x\bar U(x) = x\bar D(x) =\frac{A_U (X_U)^{-1} x^{b_U}}{\exp[(x+X_U)/\bar x] 
+ 1} +\frac{\tilde A_{U} x^{\tilde b_{U}}}{\exp(x/\bar x) + 1}~.
\label{eq6}
\end{equation}

\begin{equation}
x S(x) = x\bar S(x) = \frac{\tilde A_{U} x^{\tilde b_{U}}}{2[\exp(x/\bar x) 
+ 1]}~.
\label{eq7}
\end{equation}

\begin{equation}
x G(x) = \frac {A_G x^{b_G}} {\exp(x/\bar x)-1}~.
\label{eq8}
\end{equation}

Following the formulation developed for proton's PDFs, the $x$ distributions 
for fermions (quark and antiquark) have Fermi-Dirac functional form, while
gluon has a Bose-Einstein $x$ distribution. The two separate terms for
$U(x)$ and $\bar U(x)$ in Eqs. (\ref{eq5}) and (\ref{eq6}) refer to 
the non-diffractive 
and diffractive contribution, respectively. In the previous 
analysis~\cite{BS2019}, the diffractive term was neglected for simplicity.
As shown in the analysis of proton's PDFs in the statistical 
model~\cite{BBS2005}, the
presence of the diffractive term is important for describing the data
at the low $x$ region. Therefore, we have added the diffractive
term in this new analysis for pion's PDFs.   

A key feature of the statistical model is
that the chemical potential, $X_U$, for the quark distribution $U(x)$ becomes
$-X_U$ for the antiquark distribution $\bar U(x)$. The parameter $\bar x$
plays the role of the effective ``temperature". 
For the strange-quark distribution $S(x)$, the requirement that the $S$ and
$\bar S$ have identical $x$ distribution implies that the chemical 
potential in the
non-diffractive term must vanish. This implies that the non-diffractive
and diffractive terms for $S(x)$ have a similar functional form, and we make
the simple assumption that $S(x)$ is equal to half of the diffractive part
of  $\bar U(x)$ due to the heavier strange quark mass.

Equations (\ref{eq5})-(\ref{eq8}) contain a total of 8 parameters, namely, 
$A_U$, $X_U$, $b_U$,
$\bar x$, $\tilde A_U$, $\tilde b_U$, $A_G$, and $b_G$. In contrast, the
number of parameters for Ref.~\cite{BS2019} is significantly larger, at 14, 
even without including strange-quark sea in the pion PDFs. 

Among these 8 parameters, only 6 are truly free, due to the constraints from
two sum rules. The quark-number sum rule requires   

\begin{equation}
\int_0^1 [U(x) - \bar U(x)] dx =1~,
\label{eq:eq9}
\end{equation}
and the momentum sum rule implies

\begin{equation}
\int_0^1 x [2 U(x) + 2 \bar U(x) + 2 S(x) + G(x)] dx = 1~.
\label{eq:eq10}
\end{equation}

Since the Drell-Yan data are not sensitive to pion's gluon distribution,
it is mainly through this momentum sum rule that $G(x)$ is determined.
The strong correlation between $A_G$ and $b_G$ can be removed by requiring 

\begin{equation}
b_G = 1 + \tilde b_U~.
\label{eq:eq11}
\end{equation}
Equation~(\ref{eq:eq11}) has the interesting consequence that $G(x)$ has 
the same $x$ dependence as the diffractive part of the quark distributions
when $x \to 0$. The dominance of the gluon and sea-quark distributions at
$x \to 0$ and the strong interplay among them make Eq.~(\ref{eq:eq11})
quite a reasonable assumption. Equation~(\ref{eq:eq11}) further reduces the
number of parameters by one. 
The significant reduction in the total number of free parameters
in the statistical model, resulted from the application of symmetry 
constraints and Eq.~(\ref{eq:eq11}) discussed above, allows a stringent 
test of the statistical model for describing pion's PDFs.
It is not evident {\it a priori} that existing data can be well described by
the statistical model with very limited number of parameters.

\begin{figure}[hbp]
\begin{center}
\includegraphics[width=5.5cm]{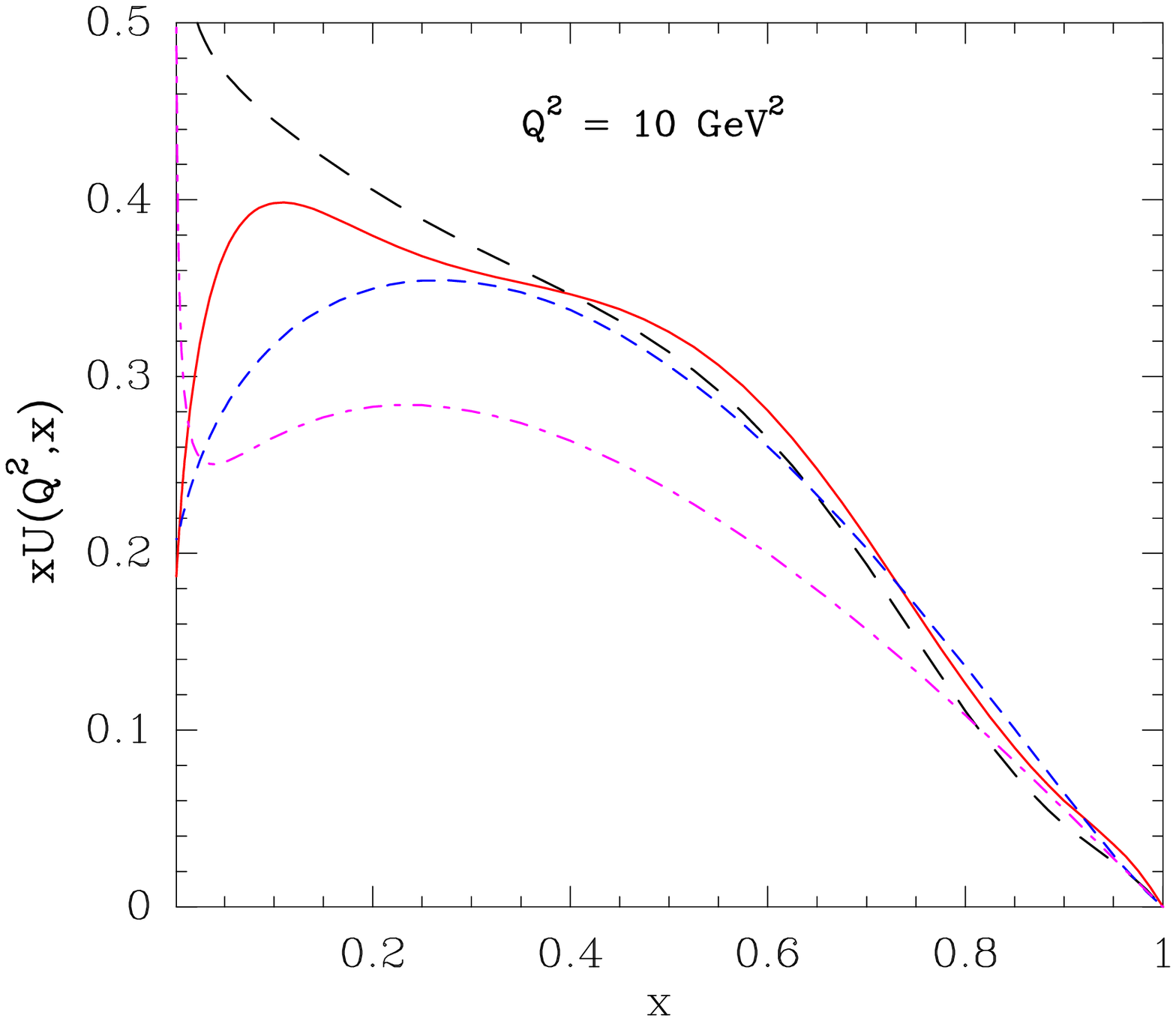}
\includegraphics[width=5.5cm]{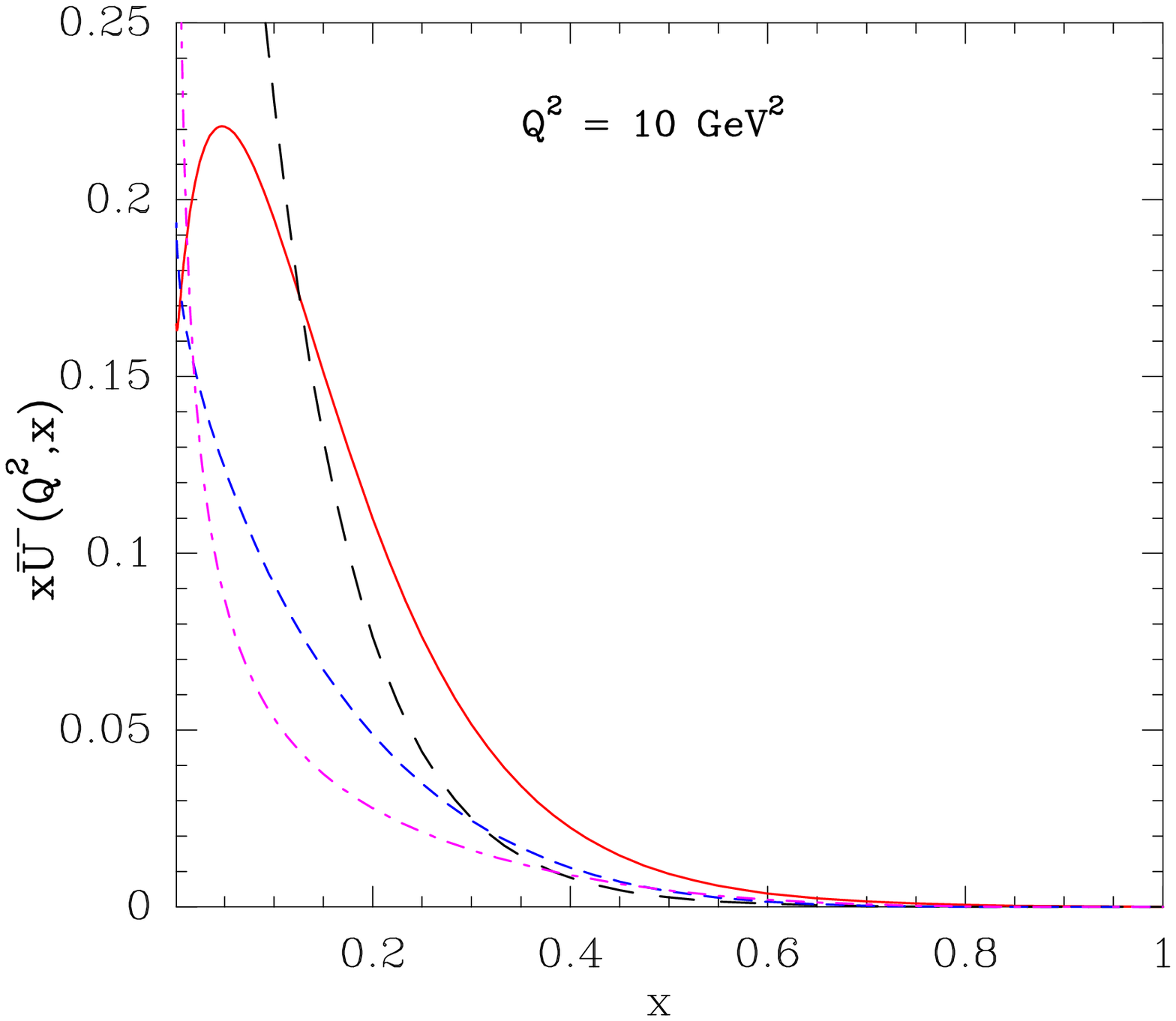}
\qquad
\includegraphics[width=5.5cm]{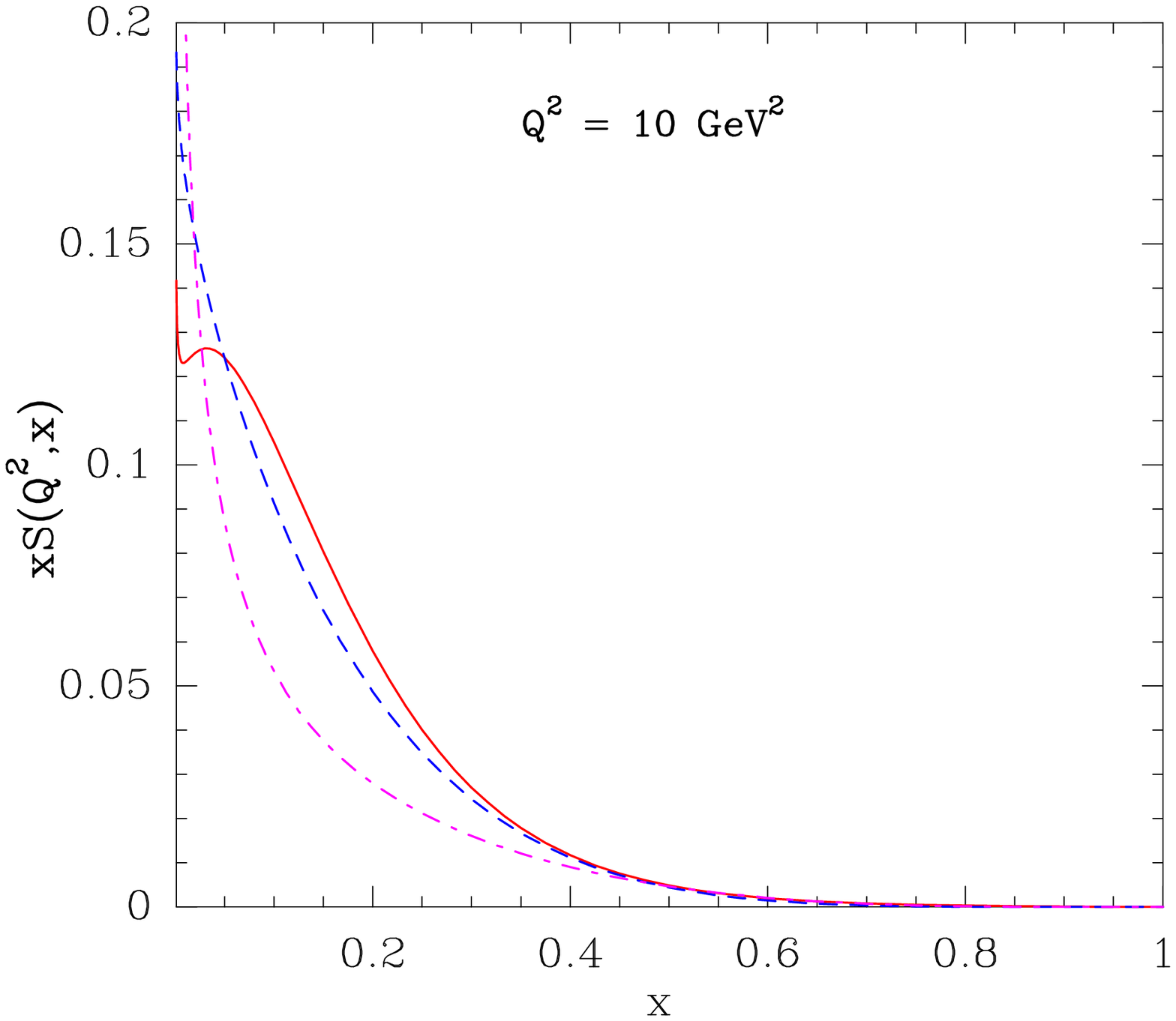}
\includegraphics[width=5.5cm]{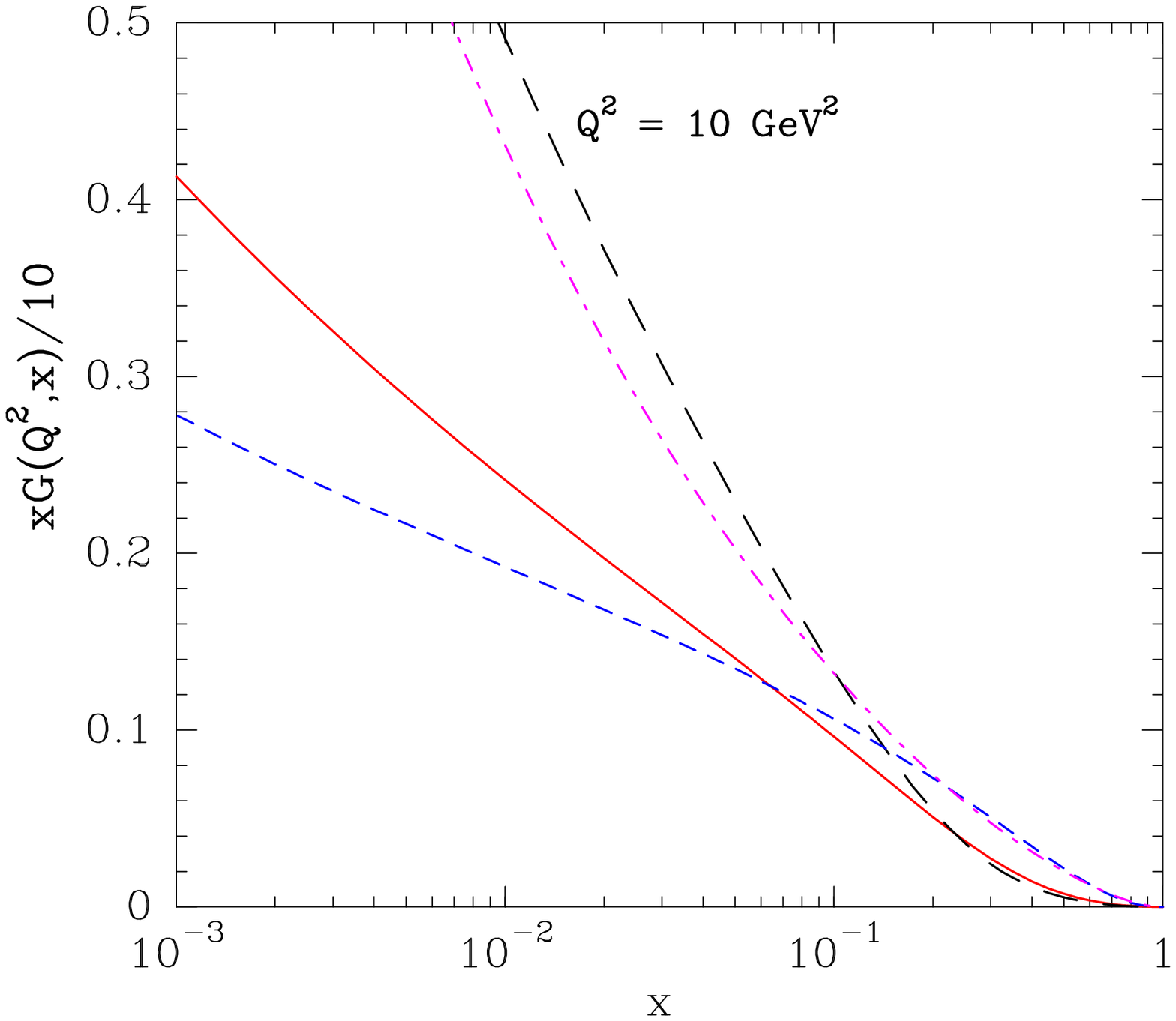}
\end{center}
\caption[*]{\baselineskip 1pt
Different $\pi^-$ parton distributions  versus $x$, after NLO QCD
evolution at $Q^2=10~\mbox{GeV}^2$. Present statistical model (solid),
previous statistical model \cite{BS2019} (long dashed),
SMRS PDFs from Ref. \cite{sutton92} (dashed),
GRV PDFs from Ref. \cite{gluck92} (dotted-dashed), are shown. 
For SMRS and GRV $x\bar S(x) = x\bar U(x)$.}
\label{PDFpi}
\end{figure}
In order to obtain the parameters for pion's PDFs according to 
the parametrizations listed in Eqs. (7) - (10), we have performed a NLO
QCD fits of $\pi^-$-induced Drell-Yan dimuon production
data on tungsten targets from E615~\cite{e615} at 252 GeV,
E326~\cite{e326} at 225 GeV, and NA10~\cite{NA10} at 194 GeV and
286 GeV. Details of the NLO calculations were described in~\cite{BS2019}. 
The QCD evolution was done using the HOPPET 
program~\cite{Salam2009}, and the $\chi^2$ minimization was performed utilizing
the CERN MINUIT program~\cite{James1994}.
\begin{figure}[hbp]
\begin{center}
\includegraphics[height=7.2cm]{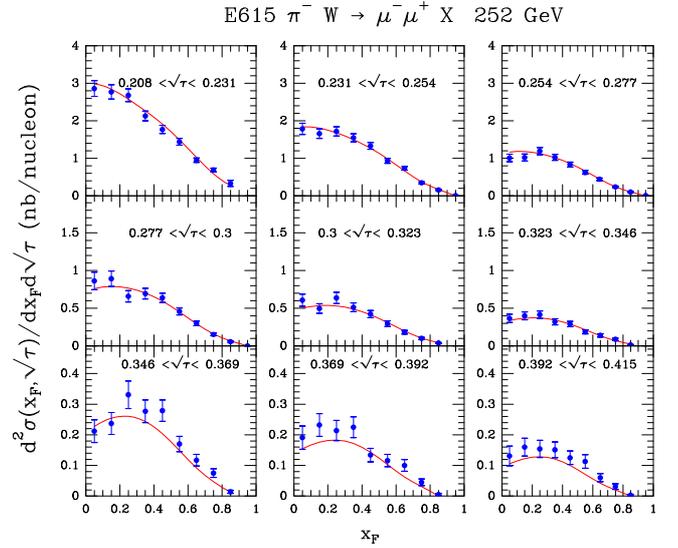}
\end{center}
\caption[*]{\baselineskip 1pt
Drell-Yan data from the E615 experiment
$\pi^- W$ at $P_{lab}^{\pi} = 252~\mbox{GeV}$ \cite{e615}.
$d^2\sigma/d\sqrt{\tau}dx_F$ versus $x_F$ for several $\sqrt{\tau}$
intervals are compared with the results of our 
global fit (solid curves).}
\label{fi1}
\end{figure}

Table 1 lists the number of data points and the values of $\chi^2$ obtained
from the best fit to these data sets. Note that the normalizations for
the absolute cross sections from various experiments contain systematic
uncertainties on the order of $\sim$ 10 percents. In the global fit,
the normalizations for various data sets are allowed to vary, as listed
in Table 1, in order to achieve improved consistency among various data 
sets.

\begin{table}[hbp]
\begin{center}
\begin{tabular}{ c c c c c }
\hline
  &P(GeV) &K  &$N_{data}$ & $\chi^2$  \\
\hline\raisebox{0pt}[12pt][6pt]
E615       & 252    & 1.066 & 91 & 123  \\[4pt]
E326    & 225    & 1.223  & 50 & 75  \\[4pt]
NA10  & 286     &   1.2928  & 23  & 8  \\[4pt]
NA10  & 194     & 1.2928  & 44  & 17  \\[4pt]
\hline\raisebox{0pt}[12pt][6pt]
Total                        & &          & 208  & 223  \\[4pt]
\hline
\end{tabular}
\caption {Results of the K factor and $\chi^2$ for each data set
from the global fit. P is the beam momentum, K the normalization
factor for the Drell-Yan cross sections, $N_{data}$ the number of 
data points.}
\label{table1}
\end{center}
\end{table}

Table~\ref{table1} shows that good $\chi^2$ values can be obtained from the 
best fits to the Drell-Yan data. The best-fit parameters, obtained at an
initial scale $Q^2_0 = 1$ GeV$^2$, are: 

\begin{align}
A_U  &=  0.80536 \pm 0.10 & b_U & = 0.5161 \pm 0.02  \nonumber \\
X_U  &=  0.7551 \pm 0.01   & \bar x & = 0.10614\pm 0.004  \nonumber \\
\tilde A_U  &= 2.2773 \pm 0.324 & \tilde b_U & = 0.4911 
\pm 0.0092 \nonumber \\
A_G  &=  31.0019 \pm 1.68 & b_G & = 1 + \tilde b_U~. 
\label{eq14}
\end{align}

It is worth noting that the temperature, $\bar x = 0.106$, found for
pion is very close
to that obtained for proton, $\bar x = 0.090$~\cite{BS2015}, indicating a 
common feature for the statistical description for baryons and mesons. 
On the other hand, the chemical potential of the valence quark for pion,
$X_U = 0.7551$, is significantly large than that for proton,
$X_U \sim 0.39$~\cite{BS2015}. This reflects the fact that baryons
contain three valence quarks while mesons only consist of two valence 
quarks.
\begin{figure}[hbp]
\begin{center}
\includegraphics[height=7.2cm]{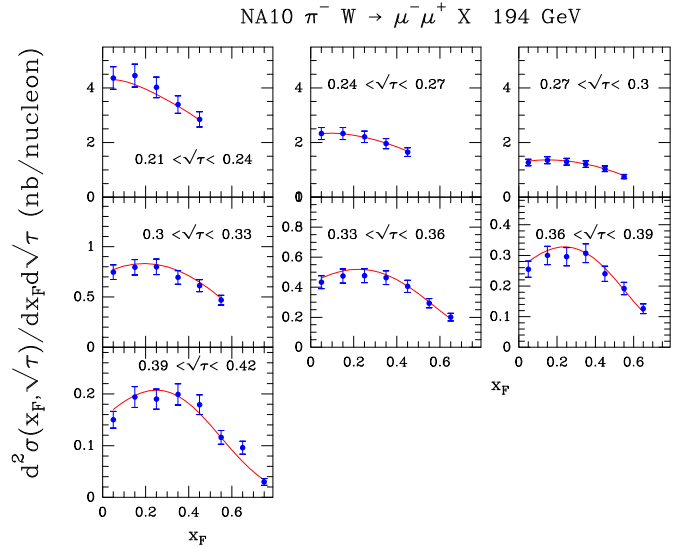}
\end{center}
\caption[*]{\baselineskip 1pt
Drell-Yan data from the NA10 experiment $\pi^- W$ at $P_{lab}^{\pi} 
= 194~\mbox{GeV}$ \cite{NA10}.  $d^2\sigma/d\sqrt{\tau}dx_F$ versus $x_F$ for
several $\sqrt{\tau}$ intervals 
are compared with the results of our
global fit (solid curves).}
\label{fi2}
\end{figure}
   
Figure~\ref{PDFpi} displays $xU(x)$, $x \bar U(x)$, 
$x S(x) = x \bar S(x)$, and $x G(x)$ at $Q^2 = 10$ GeV$^2$ obtained 
in the present 
analysis. Comparisons with the distributions from the 
previous analysis in the statistical model~\cite{BS2019} 
and global fits of
SMRS~\cite{sutton92} and GRV~\cite{gluck92} are also shown in Fig. \ref{PDFpi}.
The shape and magnitude of the pion PDFs obtained in 
the statistical model analysis are
significantly different from that of SMRS and GRV. This reflects the
very different functional forms for the PDFs in the statistical model
compared with that of the conventional global fits.   

\begin{figure}[hbp]
\begin{center}
\includegraphics[height=5.4cm]{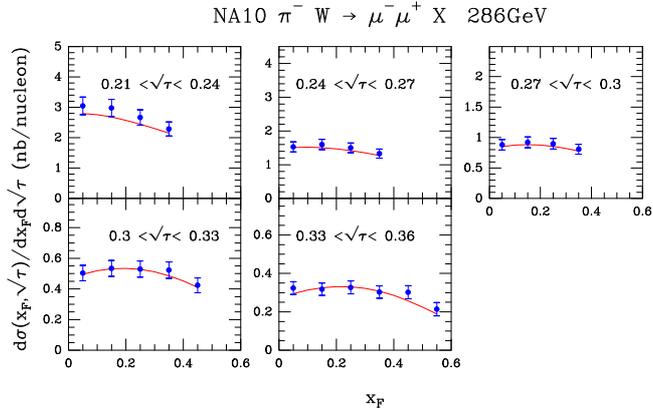}
\end{center}
\caption[*]{\baselineskip 1pt
Drell-Yan data from the NA10 experiment $\pi^- W$ at
$P_{lab}^{\pi} = 286~\mbox{GeV}$ \cite{NA10}.
$d^2\sigma/d\sqrt{\tau}dx_F$ versus $x_F$ for several
$\sqrt{\tau}$ intervals are compared with the results of our
global fit (solid curves).}
\label{fi3}
\end{figure}

In Figures~\ref{fi1} - \ref{fi3}, we show the fits to the E615 and 
NA10 data using 
the current result on pion's PDFs. 
We note that the data are well described by
the statistical model with a parametrization much simpler than the
previous analysis~\cite{BS2019}. A good agreement is also obtained for the
E326 data at 225 GeV~\cite{e326}.

\begin{figure}[hbp]
\begin{center}
\includegraphics[height=8.0cm]{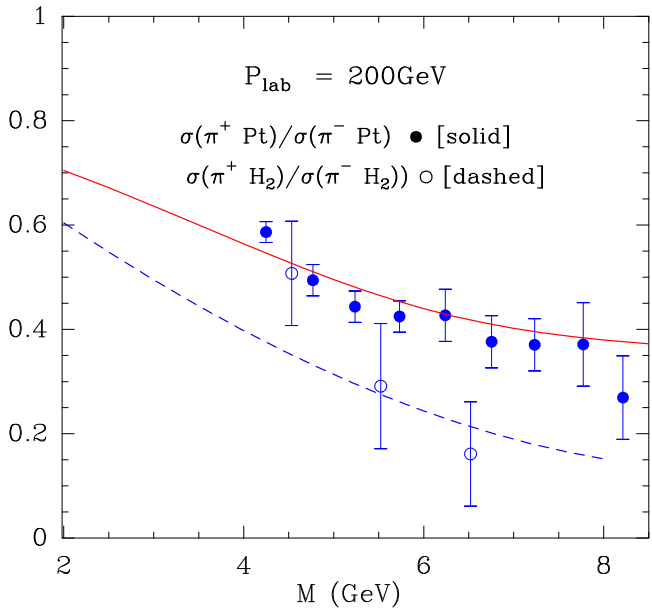}
\end{center}
\caption[*]{\baselineskip 1pt
Drell-Yan $\pi^+/\pi^-$ cross section ratio data on hydrogen and platinum 
targets at 200 GeV from the NA3 experiment~\cite{NA3}. Calculations using
the pion PDFs obtained in the current analysis are compared with the data.}
\label{fi5}
\end{figure}

To check the predictive capability of 
the current pion PDFs, we show in Figure \ref{fi5} the calculations 
for the ratios 
of $\pi^+$/$\pi^-$ Drell-Yan cross section ratios at 200 GeV on hydrogen and
platinum targets. Reasonable agreement with the NA3 data~\cite{NA3} is
found. It was suggested some time ago that a comparison of the 
$\pi^+$ and $\pi^-$ induced Drell-Yan data on an isoscalar target such as
deuteron or $^{12}$C can probe sensitively the sea-quark distribution in 
pion~\cite{londergan}. Such measurement is indeed being planned in a future
experiment at CERN~\cite{Adams}.

The Drell-Yan cross section at the NLO QCD contains the contribution
from the quark-gluon fusion process, which leads to some sensitivity 
to the gluon distribution in pion. Nevertheless, it is important to 
consider other processes which are sensitive to gluon distributions
at the LO QCD level. A recent analysis suggests that existing pion-induced 
$J/\Psi$ production can probe the gluon distribution in pion at large $x$
sensitively~\cite{Chang2020}. The prospect for including the $\pi^+$
induced Drell-Yan data as well as the pion-induced $J/\Psi$ production data
to extract the pion's PDFs in a future global fit with the statistical 
model is being considered. An extension of this approach to extract the
kaon's PDFs would also be of great interest~\cite{Peng}.

In conclusion, we have performed a new analysis to 
extract pion's PDFs in the statistical 
model using a parametrization containing fewer number of parameters than
an earlier analysis. This significant reduction in the number of parameters
is largely due to symmetry considerations. This new analysis with
reduced number of parameters shows that a good description of the 
existing $\pi^-$-induced Drell-Yan data can be obtained in the 
statistical model approach.

A comparison between the proton and pion PDFs in the statistical
model approach shows that very similar temperature parameters are found 
for both cases, suggesting the consistency of this appraoch for different
hadronic systems. The higher value of the valence-quark chemical potential 
found for pion than proton reflects the different number of valence quarks
in mesons and baryons. The predictive power of the pion PDFs obtained in
this analysis has been illustrated from the good agreement between the 
calculation and the $\pi^+/\pi^-$ Drell-Yan cross section ratio data.
New pion-induced Drell-Yan data anticipated from COMPASS would provide
further tests of the pion PDFs obtained in the statistical approach.

This statistical model approach can be extended in the future by
enlarging the data sets to include the $\pi^+$-induced Drell-Yan data as
well as the pion-induced $J/\Psi$ production data. These data further
constrain the sea-quark and gluon distributions in pion. The
prospect to extend this analysis to extract kaon's PDFs is also being
considered.

We thank Dr. Wen-Chen Chang for valuable comments. This work
was supported in part by the U.S. National Science Foundation.

\end{document}